\documentstyle{article}

\def \lleq {\lower0.9ex\hbox{ $\buildrel < \over \sim$} ~}
\def \ggeq {\lower0.9ex\hbox{ $\buildrel > \over \sim$} ~}
\def\beq{\begin{equation}}
\def\eeq{\end{equation}}
\def\ber{\begin{eqnarray}}
\def\eer{\end{eqnarray}}

\title{COSMOLOGICAL ASPECTS OF ROLLING TACHYON}
\author{M. Sami \\ Inter-University Centre for Astronomy and Astrophysics,
\\ Post Bag 4, Ganeshkhind, Pune-411 007, INDIA.\footnote{On leave from jamia Millia, New Delhi. email:sami@iucaa.ernet.in} \\ Pravabati Chingangbam and Tabish Qureshi \\ Department of Physics, 
Jamia Millia Islamia, \\New Delhi-110025, INDIA}
\begin{document}
\maketitle 
\begin{abstract}
We examine the possibility of rolling tachyon to play the dual roll of inflaton at early epochs 
and dark matter at late times. We argue that enough inflation can be generated with the rolling tachyon either by 
invoking the large number of branes or brane world assisted inflation. However, reheating is problematic in this model.
\end{abstract}


\section{Introduction}
 Cosmological inflation has become an integral part of the standard
 model of the universe. Apart from being
 capable of removing the shortcomings of the standard cosmology, it gives important clues for structure formation in the universe.
The inflationary paradigm seems to have gained a fairly good amount of
support from the recent observations on microwave background
radiation. On the other hand there have been difficulties in
obtaining accelerated expansion from fundamental theories such
as M/String theory. Recently, Sen \cite{sen1,sen2,sen3} has shown that the decay
of an unstable D-brane produces pressure-less gas 
with finite energy density that resembles classical dust. Gibbons has emphasized
the cosmological implications of  tachyonic condensate rolling towards its ground state
\cite{gibbons}, see Refs\cite{g2,kim} for further details. Rolling tachyon
matter associated with unstable D-branes has an interesting
equation of state which smoothly interpolates between -1 and
0. The tachyonic matter, therefore, might provide an explanation
for inflation at the early epochs and could contribute to some
new form of dark matter at late times \cite{shiu}, see also Refs\cite{sen4,related topics,quvedo} on the related theme and Ref\cite{matos} for an alternative
approach to rolling tachyon cosmology.
We shall review here the cosmological prospects of rolling tachyon with exponential potential. 
\section{COSMOLOGY WITH ROLLING TACHYON}
It was recently  shown by Sen that the dynamics of string tachyons in the background of an unstable D-brane can
be described by an effective field theory with Born-Infeld type action\cite{sen3}
\begin{equation}
S=\int{d^4x \sqrt{-g}\left({R \over {16\pi G}}- V(\phi) \sqrt{1+g^{\alpha \beta} \partial_\alpha \phi \partial_\beta \phi} \right)}
\label{action}
\end{equation}
where $\phi$ is the tachyon field minimally coupled to gravity. 
In a specially flat FRW cosmology the stress tensor acquires the diagonal form $T^{\alpha}_{\beta}=diag(-\rho,p,p,p)$ where the pressure and energy density are given by
\begin{equation}
\rho={V(\phi) \over {\sqrt{1-\dot{\phi}^2}}}
\end{equation}
\begin{equation}
p=-V(\phi)\sqrt{1-\dot{\phi}^2}
\end{equation}
The Friedmann equation takes the form
\begin{equation}
H^2={1 \over 3M_p^2} \rho \equiv {1 \over 3M_p^2}{V(\phi) \over {\sqrt{1-\dot{\phi}^2}}}
\label{friedman}
\end{equation}
The equation of motion of the  tachyon field which follows from (\ref{action}) is
\begin{equation}
{\ddot{\phi} \over {1-\dot{\phi}^2}}+3H \dot{\phi}+{V_{,\phi} \over V({\phi})}=0
\label{evolution eq}
\end{equation}
The conservation equation equivalent to (\ref{evolution eq}) has the usual form
$$ {\dot{\rho}_{\phi} \over \rho_{\phi}}+3H (1+\omega)=0 $$
where $\omega \equiv {p_{\phi} \over \rho_{\phi}}= \dot{\phi}^2-1$ is the
equation of state for the tachyon field. Thus a universe dominated by tachyon
field would go under accelerated expansion as long as $\dot{\phi}^2\ <\ {2
\over 3}$  which is very different from the condition of inflation for non-tachyonic
field, $\dot{\phi}^2\ <\ V(\phi)$. This is related to the fact that field potential
falls out of the equation of state in case of the tachyon field. It should also be noted
that evolution equation for tachyon field contains the logarithmic derivative of the potential.
\subsection{DYNAMICS OF TACHYONIC INFLATION IN FRW COSMOLOGY}
The tachyon potential $V(\phi) \rightarrow 0$ as $\phi \rightarrow \infty$ but its exact form
is not know at present\cite{moore}. Sen has argued that the qualitative dynamics of string theory tachyons can be described by (\ref{action}) with
 exponential potential\cite{sen3}. Padmanabhan went further to
suggest that one can construct a phenomenological runaway potential with the tachyonic equation of state
capable of leading  to a desired cosmology\cite{paddy}, see also Ref\cite{sami2} on the similar theme. In what follows we shall consider (\ref{action}) with the exponential potential
in purely phenomenological context without claiming any identification of $\phi$ with the string tachyon field.
Indeed, there are problems with inflation in case the origin of $\phi$ is traced in string theory\cite{klinde} and we will
come back to this point later.
The field equations (\ref{friedman}) and (\ref{evolution eq}) for tachyonic matter with the exponential potential
\begin{equation}
 V(\phi)=V_0e^{-\alpha \phi} 
\end{equation}
can be solved exactly in the
slow roll limit. The integration of these equations leads to\cite{sami1}

\beq
\dot{\phi}_{\it end}=\sqrt{2 \over 3},  \
\phi_{end}=-{1 \over \alpha} \ln \left( {\alpha^2 \over {6\beta^2}} \right), \
V_{end}={{\alpha^2 M_p^2} \over 2}
\eeq
where $\beta=\sqrt{ V_0/3M_p^2}$. Eq(7) is consistent with the expression of the slow roll parameter
\beq
\epsilon={M_p^2 \over 2} \left({V_{,\phi} \over V}\right)^2{1 \over V}
\eeq
The COBE normalized value for the amplitude of scalar
density perturbations 
\beq
\delta_H^2\simeq {1 \over {75 \pi^2}}{V_i^2 \over {\alpha^2 M_p^6}}\simeq 4\times  10^{-10}
\eeq
 can be used to estimate $ V_{end}$ as well as $ \alpha$. Here $ V_i$ refers to the value of the
 potential at the commencement of inflation and is related to $V_{end}$ as
\begin{equation}
V_{end}={V_i \over {2N+1}}
\end{equation}
Using Eqs. (9) and (10)
with ${\cal N} =60$ we obtain
 \beq
 V_{end}\simeq 4 \times 10^{-11}M_p^4
 \eeq
At the end of inflation, apart from the field energy density, a small amount of radiation is also
 present due to particles being produced quantum mechanically during inflation \cite{ford}
 \beq
 \rho_r = 0.01 \times g_p H_{end}^4 ~~~~~~~(10 \le g_p \le 100)
 \eeq
 which shows that the field energy density far exceeds the density in the radiation
 \beq
 {\rho_r \over{ \rho_{\phi}}} \simeq 0.01\times g_p {V_{end} \over { 9 M_p^4}}\simeq 4\times g_p \times 10^{-14}
 \eeq
 From (7) we find that $\alpha \simeq 10^{-5}M_p$ and there is no problem as long as we consider the tachyonic
model of inflation in phenomenological context. However, it would be problematic if we trace the origin of
field $\phi$ in string theory as there is no free parameter there to tune. Indeed, $\alpha$ and $V_0$
 can be expressed through string length scale and string coupling $g_s$ ~as $\alpha=\alpha_0/l_s$,~~ 
 $V_0=v_0/(2\pi)^3g_sl_s^4$ where $v_0$ and $\alpha_0$ are dimensionless constants and $V_0/v_0$ is brane tension and
$\alpha$ is the tachyon mass. Tuning $\alpha$ to $10^{-5}M_p$ leads to one of the two unacceptable situations: light mass
of the tachyon or large value of string coupling $g_s$. This problem is quite independent of the form of tachyonic potential, see the paper of  Fairbairn and Tytgat Ref\cite{shiu}. The situation can be remedied by invoking the large number of
D-branes separated by distance much larger than $l_s$\footnote{We thank S. Panda for this clarification}. The number of such branes in our case turns out to be of the order of
$10^{10}$. The other alternative could be brane assisted inflation. Indeed, the prospects of inflation in Brane World scenario
improve due to the presence of an additional quadratic density term in the
Friedmann equation. Enough inflation
can be generated in this case without tuning $\alpha$, see the paper by Bento et al\cite{shiu} and Ref \cite{sami1}.
The non-brane world  alternatives to tackle this problem are discussed by Yun-Song Piao and collaborators\cite{shiu}. \par
Regarding the late time behaviour, the  phase space analysis  of tachyon field with exponential potential was carried out in Ref \cite{
sami1}. It was shown that dust like solution is a late time attractor of the tachyonic system.
Therefore the tachyon field , in principal, could  become a candidate for  dark matter.\par
Inspite of the very attractive features of the rolling tachyon condensate, the tachyonic inflation
faces difficulties associated with reheating \cite{klinde,sami1}. A homogeneous tachyon field evolves towards its
ground state without oscillating about it and , therefore, the  conventional reheating
mechanism in tachyonic model does not work. Quantum mechanical particle
production during inflation  provides an alternative mechanism by means of
which the universe could reheat. Unfortunately, this mechanism also does not seem to
work: the  small energy density of
radiation  created in this process red-shifts faster than the energy density of the tachyon field and therefore radiation
domination in the tachyonic model of inflation never commences. However, the tachyon field
could play the  role of  dark matter if the problem associated with caustics could be overcome\cite{staro}. 
\section{Acknowledgments}
 We are thankful to  Hang Bae Kim,  Mohammad Reza Garousi and  Piao Yu-song for useful comments.


\begin{thebibliography}{99}  
\bibitem{sen1} A. Sen, arXiv: hep-th/0203211. 
\bibitem{sen2} A. Sen, arXiv: hep-th/0203265.
\bibitem{sen3} A. Sen, arXiv: hep-th/0204143.
\bibitem{gibbons} G. W. Gibbons, arXiv: hep-th/0204008.
\bibitem{g2}  G.W. Gibbons, arXiv: hep-th/0301117.
\bibitem{kim}Chanju Kim, Hang Bae Kim, Yoonbai Kim and O-Kab Kwon, hep-th/0301142.
\bibitem{shiu} 
 M. Fairbairn and M.H.G. Tytgat, arXiv: hep-th/0204070;

 A. Feinstein, arXiv: hep-th/0204140; D. Choudhury, D.Ghoshal, D. P. Jatkar and S. Panda, arXiv: hep-th/0204204; 
A. Frolov, L. Kofman and A. Starobinsky, arXiv: hep-th/0204187;
 H. S. Kim, arXiv: hep-th/0204191;
 G. Shiu and Ira Wasserman, arXiv: hep-th/0205003; T. Padmanabhan and
T. Roy Choudhury, Phys.Rev. D66 (2002) 081301[ hep-th/0205055];
 K. Hashimoto, arXiv: hep-th/0204203;
 S. Sugimoto, S. Terashima, hep-th/0205085;
 J. A. Minahan, arXiv: hep-th/0205098;
 L. Cornalba, M. S. Costa and C. Kounnas, arXiv: hep-th/0204261;
 H. B. Benaoum, arXiv: hep-th/0205140;
 xin-zhou Li, Jian-gang Hao and Dao-jun Liu, arXiv: hep-th/0204152;
 J. c. Hwang and H. Noh, arXiv: hep-th/0206100;
Y.-S. Piao, R.-G. Cai, X. Zhang, and Y.-Z. Zhang, hep-ph/0207143;
G. Shiu, S.-H. H. Tye, and I. Wasserman, hep-th/0207119;
X.-z. Li, D.-j. Liu, and J.-g. Hao, hep-th/0207146, On the tachyon inflation;
J.M. Cline, H. Firouzjahi, and P. Martineau,JHEP 0211, 041 (2002), hep-th/0207156; James M. Cline, Hassan Firouzjahi,hep-th/0301101;
Bin Wang, Elcio Abdalla, Ru-Keng Su,  hep-th/0208023;  
  S. Mukohyama, hep-th/0208094;
M.C. Bento, O. Bertolami and A.A. Sen, hep-th/0208124;
J.-g. Hao and X.-z. Li, Phys. Rev. D66, 087301 (2002), hep-th/0209041;
Chanju Kim , Hang Bae Kim and Yoonbai Kim, hep-th/021010.
J.S. Bagla, H.K. Jassal, and T. Padmanabhan, astro-ph/0212198;
Yun-Song Piao, Qing-Guo Huang, Xinmin Zhang, Yuan-Zhong Zhang, hep-ph/0212219.

\bibitem{sen4} Ashoke Sen, hep-th/0207105; Partha Mukhopadhyay and Ashoke Sen,  hep-th/0208142;  Ashoke Sen,  hep-th/0209122.
\bibitem{related topics} Akira Ishida, Shozo Uehara, hep-th/0206102 ;T. Mehen and Brian Wecht, hep-th/0206212;
Kazutoshi Ohta, Takashi Yokono,hep-th/0207004; Nicolas Moeller, Barton Zwiebach, hep-th/0207107; T. Okuda and S. Sugimoto,
hep-th/0208196;  Gary Gibbons, Koji Hashimoto, Piljin Yi, hep-th/0209034; Mohammad R. Garousi, hep-th/0003122; Mohammad R. Garousi,hep-th/0209068;
J. Kluson, hep-th/0209255; Sayan Kar, hep-th/0210108; Haewon Lee, W. S. l'Yi, hep-th/0210221; Kenji Hotta,  hep-th/0212063;
Soo-Jong Rey, Shigeki Sugimoto,  hep-th/0301049;  Chanju Kim, Hang Bae Kim, Yoonbai Kim, O-Kab Kwon, hep-th/0301076; 
Akira Ishida and Shozo Uehara, hep-th/0301179.
\bibitem{quvedo} M. Gasperini, G. Veneziano, The Pre-Big Bang Scenario in String Cosmology,  hep-th/0207130; F. Quevedo, Lectures 
on Strings/Brane Cosmology, hep-th/0210292.
\bibitem{matos}  G.A. Diamandis, B.C. Georgalas , N.E. Mavromatos, E. Papantonopoulos,  hep-th/0203241;
 G.A. Diamandis, B.C. Georgalas , N.E. Mavromatos,
 E. Papantonopoulos, I. Pappa,  hep-th/0107124.

\bibitem{moore} A. A. Gerasimov, S. L. Shatashvili, JHEP 0010 034(2000) [hep-th/0009103]; A. Minahan and B. Zwiebach;
JHEP 0103 038 (2001) [hep-th/0009246]; D. Kutasov,
A. Tseytlin, J. Math Phys., 42 2854 (2001).
\bibitem{paddy} T. Padmanabhan, Phys.Rev. D66 (2002) 021301[hep-th/0204150].
\bibitem{sami2} M. Sami, hep-th/0205146.
\bibitem{klinde} L. Kofman and A. Linde, Phys.Lett. B545 (2002) 8-16[hep-th/0205121].
\bibitem{sami1} M. Sami, P. Chingangbam and T. Qureshi,Phys.Rev. D66 (2002) 043530[hep-th/0205179].
\bibitem{ford} L.H. Ford, Phys. Rev. D, {\bf 35}, 2955 (1987), B. Spokoiny, Phys. Lett. {\bf B
315}, 40 (1993).



\bibitem{papa} E. Papantonopoulos and Papa, Mod. Phys. Lett. {\bf A15}, 2145 (2000) [hep-th/0001183] ; Phys. Rev. {\bf D63}
            , 103506 (2000) [hep-th/0010014] ;
            S. H. S. Alexander, Phys. Rev. {\bf D65},023507 [ hep-th/0105032];
            A. Mazumdar, S. Panda, A. Perez-Lorenzana, Nucl. Phys. {\bf B614}, 101, (2001)[hep-th-0107058];
            Shinji Mukohyama, arXiv: hep-th/0204084.
\bibitem{staro} Gary Felder, Lev Kofman and Alexei Starobinsky ,JHEP 0209 (2002) 026[hep-th/0208019].




\end{thebibliography}
\end{document}